\documentclass[10pt,superscriptaddress,twocolumn,amsmath,amssymb,aps,prb,longbibliography,reprint]{revtex4-2}

\usepackage{mathrsfs}
\usepackage{graphicx}
\usepackage{dcolumn}
\usepackage{bm}
\usepackage{color}

\usepackage{multirow}
\usepackage[colorlinks,bookmarks=false,citecolor=blue,linkcolor=red,urlcolor=blue]{hyperref}

\newcommand{\be}{\begin{equation}}
\newcommand{\ee}{\end{equation}}
\newcommand{\bea}{\begin{eqnarray}}
\newcommand{\eea}{\end{eqnarray}}

\definecolor{purple}{RGB}{128,0,128}

\begin{document}
\title{The $s\pm$ pairing symmetry from electron-phonon coupling in La$_3$Ni$_2$O$_7$ under pressure}

\author{Yucong Yin}
\affiliation{Institute of Theoretical Physics, School of Physics and Optoelectronic Engineering,
Beijing University of Technology, Beijing, 100124, China}

\author{Jun Zhan}
\affiliation{Beijing National Laboratory for Condensed Matter Physics and Institute of Physics, Chinese Academy of Sciences, Beijing 100190, China}
\affiliation{School of Physical Sciences, University of Chinese Academy of Sciences, Beijing 100190, China}

 \author{Boyang Liu}\email{boyangleo@gmail.com}
\affiliation{Institute of Theoretical Physics, School of Physics and Optoelectronic Engineering,
Beijing University of Technology, Beijing, 100124, China}
\author{Xinloong Han}
\email{hanxinloong@gmail.com}
\affiliation{Department of Physics, Capital Normal University, Beijing 100048, China}
\affiliation{Kavli Institute for Theoretical Sciences, University of Chinese Academy of Sciences, Beijing 100190, China}

\date{\today}
\begin{abstract}

The recently discovered bilayer Ruddlesden-Popper nickelate La$_3$Ni$_2$O$_7$ exhibits superconductivity with a remarkable transition temperature $T_c\approx 80 $ K under applied pressures above 14.0 GPa. This discovery of new family of high-temperature superconductors has garnered significant attention in the condensed matter physics community. In this work, we assume the this high-temperature superconductor is mediated by phonons and investigate the pairing symmetry in two distinct models: (i) the full-coupling case, where the Ni-$d_{x^2-y^2}$ and Ni-$d_{3z^2-r^2}$ orbitals are treated equally in both interlayer and intralayer coupling interactions, and (ii) the half-coupling case, where the intralayer coupling involves only the $d_{x^2-y^2}$ orbital, while the interlayer coupling is restricted to the $d_{3z^2-r^2}$ orbital. Our calculations reveal that the interlayer coupling favors an $s\pm$-wave superconducting state, whereas the intralayer coupling promotes an $s++$-wave symmetry. Additionally, we discuss the implications of pair-hopping interactions on the superconducting properties. These findings provide valuable insights into the pairing mechanisms and symmetry of this newly discovered high-temperature superconductor.

% The newly discovered bilayer Ruddlesden-Popper nickelate La$_3$Ni$_2$O$_7$ establishes a superconducting properties with an astonishing transition temperature $T_c\approx80$ K at pressures between 14.0 GPa and 43.5 GPa. The discovery of a new family of high temperature (T$_c$) materials has highly attracted many attentions. In this work, we investigate the pairing symmetry of this high T$_c$ superconductor by considering two models, namely the full-coupling case where Ni-$d_{x^2-y^2}$ and Ni-$d_{z^2}$ orbitals are treated equally in the interlayer and intralayer coupling interactions, and half-coupling case in which the intralayer coupling only contains the $d_{x^2-y^2}$ orbital and interlayer coupling only contains $d_{3z^2-r^2}$ orbital. In these two models, phonon plays the role of pairing glue. Our calculations in the two models uncover that the interlayer coupling tends to trigger the $s\pm$ superconducting state, while the intralayer coupling will lead to the $s++$ superconductivity. Moreover, we also discuss the consequence of the pair hopping interactions. 

\end{abstract}
 \maketitle

\section{Introduction}

The search for a new family of high transition temperature (high-$T_c$) superconductors remains the central task in the condensed matter community, since the discovery of copper-based high-$T_c$ superconductors\cite{bednorz1986possible,anderson1987resonating,lee2006doping,keimer2015quantum}. Drawing inspiration from the crystal and electronic structures of cuprates, the iron-based superconductors\cite{kamihara2008iron,Daghofer2008model,Raghu2008minimal,chubukov2008Magnetism,Eschrig2009tight,Stewart2011Superconductivity} and the recent infinite layer nickelate superconductors\cite{li2019superconductivity,Been2021prx,Lee2004prb,hepting2020electronic,Botana2020prx,wu2020prb,jiang2020prl} are discovered.  Notably, the discovery of superconducting property in bilayer Ruddlesden-Popper nickelate La$_3$Ni$_2$O$_7$ (LNO)\cite{sun2023signatures}, which exhibits a remarkable transition temperature $T_c\approx 80$ K under high pressure exceeding 14 GPa, has sparked intensive investigation into this novel superconducting material\cite{,hou2307emergence,zhang2024high,PhysRevX.14.011040,luo2023bilayer,lechermann2023electronic,gu2023effective,yang2023possible,christiansson2023correlated,sakakibara2024possible,shilenko2023correlated,shen2023effective,liu2023electronic,wu2023charge,chen2023critical,cao2024flat,liu2023s,zhang2306electronic,lu2024interlayer,zhang2024structural,Hanbit2023typeII,liao2023electron,qu2024bilayer,yang2023minimal,jiang2024high,zhang2023trends,huang2023impurity,tian2024correlation,qin2023high,heier2024competing,khasanov2024pressure,meng2024density,yang2024orbital,Ku2024PRL,yi2024nature,liu2023evidence,geisler2024structural,xia2025sensitive,Jiang2025PRL,zhou2025ambient,dong2024visualization,dan2024spin}. Subsequent studies\cite{hou2307emergence,zhang2024high,PhysRevX.14.011040} have further confirmed the observation of zero electrical resistance, solidifying LNO as a promising candidate for further exploration in the field of high-$T_c$ superconductivity. Increasing pressure causes the structure to undergo a transition from the orthorhombic structure of the Amam space group to a regular AA-stacking structure of the Fmmm space group. In contrast to the $d^9$ electronic configuration in cuprates, the normal valence band in LNO manifests the $d^{7.5}$ configuration\cite{sun2023signatures,zhang2024high}. The density functional theory (DFT) further reveals that the band structure in the vicinity of the Fermi level under pressure is mainly dominated by the Ni-$d_{x^2-y^2}$ and Ni-$d_{3z^2-r^2}$ orbitals\cite{sun2023signatures}.  These different features imply the fundamental diverse pairing symmetry from the cuprates.  Interestingly, in addition to superconductivity, the charge/spin density wave, and non-Fermi-liquid behavior are also suggested by transport measurements\cite{sun2023signatures,liu2023evidence,zhang2024high,liu2023electronic}.

A fundamental question in the study of this new family of high-$T_c$ superconductors revolves around elucidating the underlying superconducting mechanism. Current theoretical efforts have predominantly focused on understanding superconductivity in terms of the unique band structure and electronic correlations present in this material. However, recent experimental findings suggest that electron-phonon coupling may play a pivotal role in determining the electronic properties. The ongoing debate among several research groups on the strength of electron-phonon coupling\cite{talantsev2024debye,li2024ultrafast,you2025unlikelihood,ouyang2024absence,zhan2024cooperation} (EPC) raises a critical question: What would be the pairing symmetry in pressurized LNO if the superconducting state is indeed mediated by electron-phonon ($e$-ph) coupling? 

To address this question, we present a systematic investigation of the superconducting pairing symmetry within the framework of BCS theory by solving the linearized gap equations in momentum space. Specifically, we introduce and analyze two distinct models: (i) the full-coupling case, where the Ni-$d_{x^2-y^2}$ and Ni-$d_{3z^2-r^2}$ orbitals are treated on an equal footing in the interactions, and (ii) the half-coupling case, where the intralayer coupling involves only the $d_{x^2-y^2}$ orbital, while the interlayer coupling is restricted to the $d_{3z^2-r^2}$ orbital. Our calculations reveal that the interlayer coupling favors a spin-singlet $s\pm$ pairing symmetry, whereas the intralayer coupling competes by stabilizing a spin-singlet $s++$ pairing symmetry.

 The reminder of this paper is organized as follows. In the section \ref{sec:model_and_method}, we introduce our two BCS models band model and demonstrate the method to investigate the superconducting pairing symmetry. In the section \ref{sec:superconducting_state_for_the_full_coupling_case} and \ref{sec:pairing_symmetry_in_half_coupling_case}, we present the numerical results in the full-coupling case and in the half-coupling case, respectively. In section \ref{sec:the_role_of_pair_hopping_interaction}, we further discuss the role of pair hopping term in determining the pairing symmetry. Finally, in the section \ref{sec:discussions_and_conclusions} we give our further discussions and conclude this paper.

 %In the single band superconductors, the massless phase  or Bogouliubov-Anderson-Goldstone mode\cite{Anderson1958,Bogolyubov_1959} (BAG) as the consequence of $U(1)$ spontaneous symmetry breaking, and massive Higgs mode associated with the fluctuation of amplitude of superconducting order parameter emerge. Despite these two modes, the fluctuation of relative phase can intrigue the Leggett mode\cite{leggett1966number} in multiband superconductors, with the mass proportional to the interband coupling\cite{leggett1966number,xiaohu2012massless_prl}. Usually, the massive Leggett mode decays by breaking quasiparticles. However, it is possible that the Leggett mode becomes massless such that it become stable in the TRSB superconducting phase. Moreover, Higgs modes and phase modes couple together in the TRSB phase, modifying their masses.

\section{Model and method}\label{sec:model_and_method}
\begin{figure}
	\includegraphics[width=0.45\textwidth]{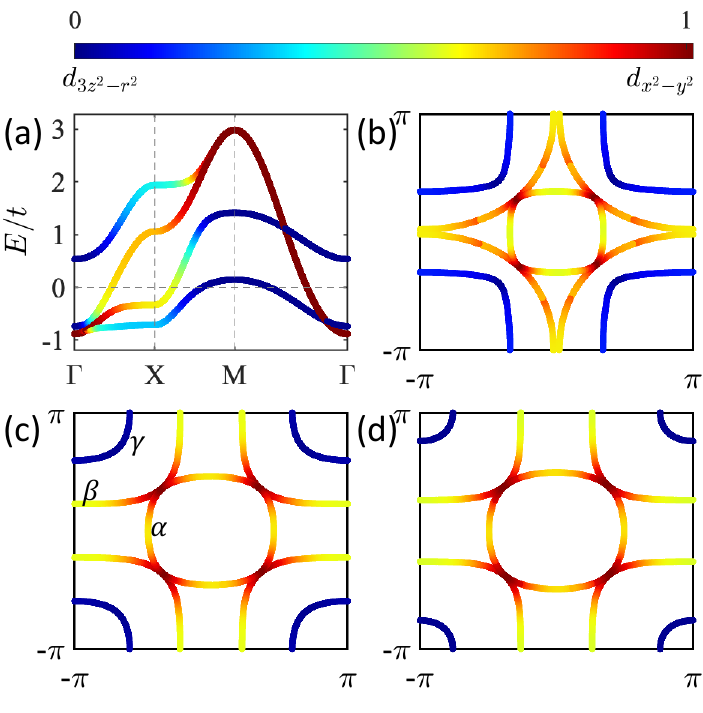}
	\caption{The orbital-weight band structure\cite{luo2023bilayer} (a) and Fermi surfaces (b-d) of the bilayer tight-binding model in the pressured LNO. (b) labels the Fermi surface with $\mu=-0.328eV$ where it manifest a van Hove singularity. (c) and (d) illustrates the Fermi surface with $\mu=0eV$ and $0.1eV$, respectively.}
	\label{fig:Fig1}
\end{figure}

The low-energy electronic structure of bilayer La$_3$Ni$_2$O$_7$, as revealed by DFT calculations and experiments, is predominantly governed by the Ni-$d_{x^2-y^2}$ and $d_{3z^2-r^2}$ orbitals\cite{sun2023signatures,luo2023bilayer,lechermann2023electronic,gu2023effective}. Motivated by this orbital-selective character, we take a minimal bilayer two-orbital model adapted from Ref. \cite{luo2023bilayer}. This effective model successfully encapsulates the essential features of the low-energy band structure near the Fermi level, including orbital hybridization effects and interlayer coupling mechanisms. The model reads
% The low energy of electronic structure of the bilayer La$_3$Ni$_2$O$_7$ is dominated by Ni-$d_{x^2-y^2}$ and $d_{3z^2-r^2}$ orbitals from DFT calculations. Hence in this paper, we take the bilayer two-orbital model from reference\cite{luo2023bilayer} which can capture the key ingredents of the low-energy electronic band structure,
\bea
H_0=\sum_{{\bf k}\sigma}\Psi^{\dagger}_{{\bf k}\sigma}\big[H_0({\bf k})-\mu\big]\Psi_{{\bf k}\sigma},
\eea 
where the basis is defined as $\Psi_{\sigma}=\left(c_{tx^2\sigma},c_{tz^2\sigma},c_{bx^2\sigma},c_{bz^2\sigma}\right)^{T}$ with field operator $c_{\eta\alpha\sigma}$ denotes the annihilation of an electron  with spin $\sigma=\uparrow,\downarrow$
in the top ($\eta=t$) or bottom ($\eta=b$) layer. The orbital index $x^2$ ($z^2$) labels the Ni-$d_{x^2-y^2}$ ($d_{3z^2-r^2}$) orbital. The chemical potential is denoted by $\mu$. The matrix $H_0$ is written as 
\bea
H_0({\bf k})=\begin{pmatrix}
T_{{\bf k}}^{x^2} & V_{{\bf k}} & t^{x^2}_{\perp} & V_{\bf k}^{\prime} \\
V_{{\bf k}} & T_{{\bf k}}^{z^2} & V_{\bf k}^{\prime} &  t^{z^2}_{\perp} \\
t^{x^2}_{\perp} & V_{\bf k}^{\prime} & T_{{\bf k}}^{x^2} &V_{{\bf k}} \\
V_{\bf k}^{\prime} & t^{z^2}_{\perp} & V_{{\bf k}} & T_{{\bf k}}^{z^2}
\end{pmatrix}
\eea
with 
\bea
&T_{{\bf k}}^{\eta}=2t^{\eta}_1(\cos k_x+\cos k_y)+4t^{\eta}_2\cos k_x\cos k_y+\epsilon^{\eta},\\
&V_{\bf k}=2t_3(\cos k_x-\cos k_y), V^{\prime}_{\bf k}=2t_4(\cos k_x-\cos k_y).
\eea
Here $T^{\eta=x^2,z^{2}}$ represents the intralayer intraorbital hopping term, while $V_{\bf k}$ ($V^{\prime}_{\bf k}$) demonstrates the interorbital intralayer (interlayer) hopping. $t_{\perp}^{x^2}$ ($t_{\perp}^{z^2}$) denotes the interlayer hopping term in Ni-$d_{x^2-y^2}$ ($d_{3z^2-r^2}$) orbital. The detailed values of hopping parameters $t_1^\eta$, $t_2^\eta$, $t_3$, $t_4$ can be found in the reference\cite{luo2023bilayer}. Fig. \ref{fig:Fig1}(a) presents the calculated band structures of the system, while panels (b)-(d) display the corresponding Fermi surfaces at three representative chemical potentials. In addition to the prominent high-$T_c$ superconducting regime near the critical doping level reported in Ref. \cite{sun2023signatures}, we systematically investigate two characteristic doping scenarios: (i) a hole-doped regime with $\mu=-0.328eV$, where the Fermi surface topology exhibits a van Hove singularity, and (ii) an electron-doped regime with $\mu=0.1eV$. Fig. \ref{fig:Fig1}(b-d) reveals that there exist three Fermi surfaces: two hole-like pockets $\beta$ and $\gamma$ originating from the interlayer antibonding states of the Ni-$d_{x^2-y^2}$ and bonding state of $d_{3z^2-r^2}$ orbitals; one electron-like pocket formed by the interlayer bonding states of both $d_{x^2-y^2}$ and $d_{3z^2-r^2}$ orbitals. Notably, the orbital-projection analysis demonstrates that the $\gamma$ pocket predominantly derives its spectral weight from the $d_{3z^2-r^2}$ orbital character.
% In Fig. \ref{fig:Fig1}, we plot the band structures as shown in (a), and Fermi surfaces with different chemical potentials as illustrated in (b-d). Here we consider Despite the most astonishing near point where the high-$T_c$ superconducting state occurs\cite{sun2023signatures}, we also consider two doping cases near the neutral point: (i) the hole doping case with $\mu=-0.328$ where the point manifests a van Hove singularity, and (ii) electron doping case with $\mu=0.1$. From Fig. \ref{fig:Fig1} (b-d), one can notice that there exist three Fermi surfaces: two hole pockets labeled as $\beta$ and $\gamma$, arising from the interlayer antibonding state of $d_{x^2-y^2}$ and bonding state of $d_{3z^2-r^2}$; one electron pocket from interlayer bonding state of $d_{x^2-y^2}$ and $d_{3z^2-r^2}$. Furthermore, Fig. \ref{fig:Fig1} also demonstrates the $\gamma$ pocket is mainly contributed by the $d_{3z^2-r^2}$. 

To elucidate the superconductivity mediated by $e$-ph coupling in the pressured LNO, we implement two distinct Bardeen-Cooper-Schrieffer (BCS) effective pairing interactions. The first one is the full-coupling case which reads
\begin{align}
&H_{I}^{f}=-g_{1}^{f}\sum_{\eta \alpha s s^{\prime};{\bf k},{\bf k}^\prime}\left[c_{\eta\alpha s}^{\dagger}(\boldsymbol{k})c_{\eta\alpha s^{\prime}}^{\dagger}(-\boldsymbol{k}) c_{\eta\alpha s^{\prime}}(-\boldsymbol{k}^{\prime})c_{\eta\alpha s}(\boldsymbol{k}^{\prime})\right] \nonumber \\
&-g_{2}^{f}\sum_{\alpha s s^{\prime};{\bf k},{\bf k}^\prime}\left[c_{t\alpha s}^{\dagger}(\boldsymbol{k})c_{b\alpha s^{\prime}}^{\dagger}(-\boldsymbol{k})c_{b\alpha s^{\prime}}(-\boldsymbol{k}^{\prime})c_{t\alpha s}(\boldsymbol{k}^{\prime})\right],\label{Eq:full}
\end{align}
% \\ 
% -g_{3}^{f}\sum_{s s^{\prime};{\bf k},{\bf k}^\prime}\left[c_{\eta x^2s}^{\dagger}(\boldsymbol{k})c_{\eta x^2s^{\prime}}^{\dagger}(-\boldsymbol{k})c_{\eta z^2s^{\prime}}(-\boldsymbol{k}^{\prime})c_{\eta z^2s}(\boldsymbol{k}^{\prime})\right],
in which we take two orbital on equal footing when considering the intralayer or interlayer interactions. The parameters $g_{1}^{f}$ and $g_{2}^f$ represents the intralayer and interlayer pairing strength within the same orbital, respectively, for the full-coupling case. 
% $g_P^f$ is the pairing hopping term between two orbitals. 
The second type of interacting Hamiltonian is written as
\begin{align}
&H_{I}^{h}=-g_{1}^{h}\sum_{\eta s s^{\prime};{\bf k},{\bf k}^\prime}\left[c_{\eta x^2 s}^{\dagger}(\boldsymbol{k})c_{\eta x^2 s^{\prime}}^{\dagger}(-\boldsymbol{k}) c_{\eta x^2 s^{\prime}}(-\boldsymbol{k}^{\prime})c_{\eta x^2 s}(\boldsymbol{k}^{\prime})\right] \nonumber\\
&-g_{2}^{h}\sum_{s s^{\prime};{\bf k},{\bf k}^\prime}\left[c_{t z^2 s}^{\dagger}(\boldsymbol{k})c_{bz^2 s^{\prime}}^{\dagger}(-\boldsymbol{k})c_{bz^2 s^{\prime}}(-\boldsymbol{k}^{\prime})c_{t z^2 s}(\boldsymbol{k}^{\prime})\right],\label{Eq:half}
\end{align}
\begin{figure}
	\includegraphics[width=0.5\textwidth]{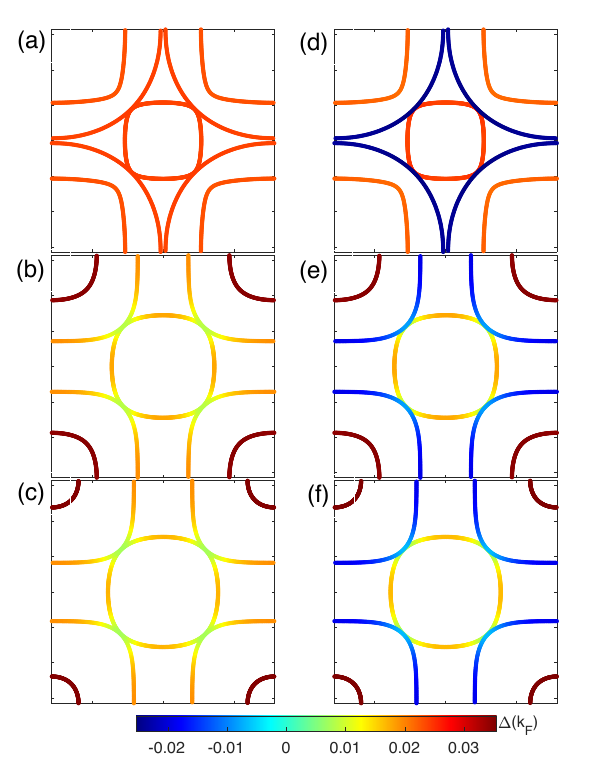}
	\caption{The pairing gap in the full-coupling case with different parameters. (a), (b) and 
 (c) illustrates the numerical result with only intralayer coupling $g_2^f=0.12$ for $\mu=-0.328eV$, $0eV$ and $0.1eV$ respectively. (d), (e) and (f) illustrates the numerical result with only interlayer coupling $g_1^f=0.12$ for $\mu=-0.328eV$, $0eV$ and $0.1eV$ respectively.}
	\label{fig:Fig2}
\end{figure}which is referred to as the half-coupling case, where we only consider the intralayer BCS pairing interaction occurring in the $d_{x^2-y^2}$ orbital, while the interlayer BCS pairing interaction occurs in the $d_{3z^2-r^2}$ orbital. The parameter $g_1^h$ ($g_2^h$) is the intralayer (interlayer) interaction strength. The intralayer interaction in this model is attributed to the coupling of the $B_{1g}$ phonon mode, which involves the in-plane motion of oxygen atoms, to the $d_{x^2-y^2}$ orbital. In contrast, the interlayer interaction is associated with two $A_{1g}$ phonon modes, which are strongly coupled to the $d_{3z^2-r^2}$ orbital.
% The intralayer interaction in the model can be attributed to the fact that the $B_{1g}$ phonon mode involving the in-plane motion of oxygen atom is predominantly coupled to the $d_{x^2-y^2}$ orbital.
% In constrast, the interlayer interaction is associated with two $A_{1g}$ phonon modes which is strongly coupled to the $d_{3z^2-r^2}$ orbital.
The the low-energy Hamiltonian $H$ is $H=H_0+H_{I}^{f,h}$. 
The orbital and layer basis $c_{\eta \alpha s}({\bf k})$ can be mapped to the band basis $c_{b s}({\bf k})$ by introducing a unitary matrix as
\bea
c_{\eta \alpha s}({\bf k})=\sum_{\beta}U_{\eta \alpha \beta}({\bf k})c_{\beta s}({\bf k})
\eea
where $\beta$ is the band index. Here we only consider the intraband pairing, thus the projected interacting Hamiltonian is 
\bea
&&H_{BCS}=-\sum_{{\bf k},{\bf k}^\prime}\sum_{bb^\prime;ss^{\prime}}g^{\beta,\beta^\prime}({\bf k},{\bf k}^\prime) c_{\beta s}^\dagger({\bf k})c_{\beta s^\prime}^\dagger(-{\bf k})\nonumber \\
&&\times c_{\beta^\prime s^\prime}(-{\bf k}^\prime)c_{\beta^\prime s}({\bf k}^\prime)\label{Eq:BCS}
\eea
and we define $g^{\beta,\beta^\prime}({\bf k},{\bf k}^{\prime})=g_1^{\beta,\beta^\prime}({\bf k},{\bf k}^{\prime})+g_2^{\beta,\beta^\prime}({\bf k},{\bf k}^{\prime})$ where $g_1^{\beta,\beta^\prime}$ and $g_2^{\beta,\beta^\prime}$ comes from the contribution of the intralayer and interlayer, respectively. Within the mean-field approximation, we introduce the order parameters $\Delta^{\beta}_{ss^\prime}({\bf k})$ by the following
\bea
\Delta^{(\beta)}_{ss^\prime}({\bf k})\equiv \sum_{\beta^{\prime}}\sum_{{\bf k}^{\prime}} g^{\beta,\beta^\prime}({\bf k},{\bf k}^\prime)\langle c_{\beta^\prime s^\prime}(-{\bf k}^\prime)c_{\beta^\prime}({\bf k}^\prime)\rangle.\label{Eq:gap}
\eea\begin{figure}
	\includegraphics[width=0.45\textwidth]{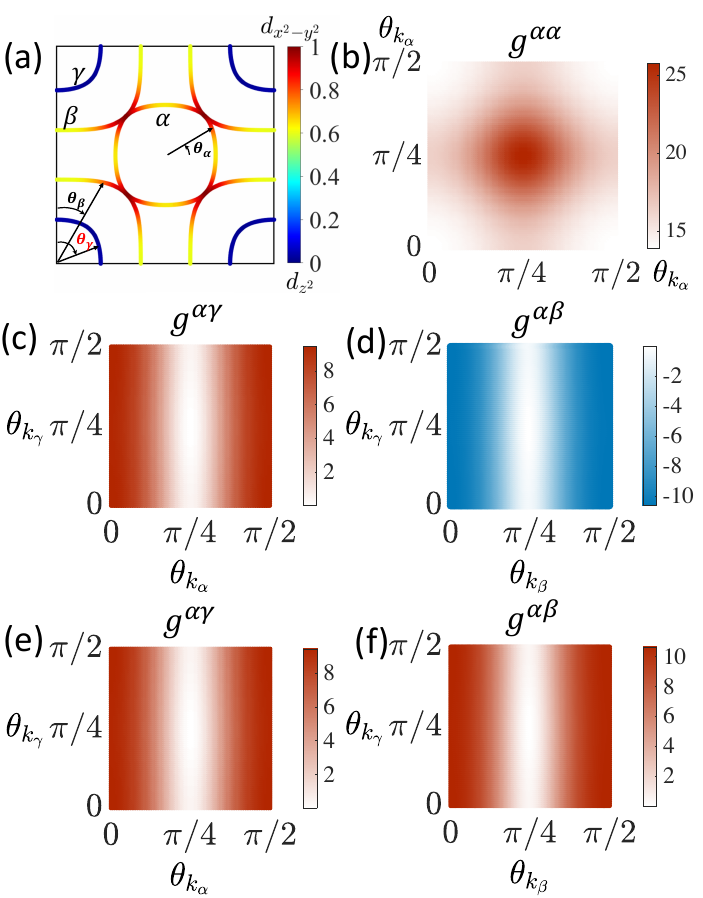}
	\caption{(a)The momentum distribution at Fermi surfaces when calculating effective interacting strength with fixed $\mu=0eV$. Effective interacting strength $g^{\alpha\alpha}$ (b), $g^{\alpha\gamma}$ (c) and $g^{\alpha\beta}$ (d) obtained in the full-coupling case with the pure interlayer coupling $g_2^f=0.12$. Effective interacting strength $g^{\alpha\gamma}$ (e) and $g^{\alpha\beta}$ (f) obtained in the full-coupling case with the pure intralayer coupling $g_1^f=0.06$. }
	\label{fig:Fig3}
\end{figure}
To get the gap equation, we introduce the Bogoliubov quasiparticles $a_{\beta}({\bf k})$ and $a_{\beta}^{\dagger}(-{\bf k})$ and they can be connected to original creation and annihilation operators $c_{\beta}({\bf k})$ by the following Bogoliubov transformation 
\bea
&&c_{\beta \uparrow}({\bf k})=u_{\beta{\bf k}} a_{\beta} ({\bf k})+v_{\beta{\bf k}} a^{\dagger}_{\beta} (-{\bf k}),\label{Eq:Bogolibov1}\\ 
&&c_{\beta \downarrow}^{\dagger}(-{\bf k})=-v_{\beta{\bf k}} a_{\beta} (-{\bf k})+u_{\beta{\bf k}} a^{\dagger}_{\beta} ({\bf k}),
\eea
with parameters
\bea
u^2_{\beta {\bf k}}=\frac{1}{2}\left(1+\frac{\varepsilon_\beta ({\bf k})}{E_\beta({\bf k})}\right),v^2_{\beta {\bf k}}=\frac{1}{2}\left(1-\frac{\varepsilon_\beta ({\bf k})}{E_\beta({\bf k})}\right),\label{Eq:uvvk}
\eea
where $\varepsilon_\beta({\bf k})$ denotes $\beta$-band's energy and $E_{\beta}({\bf k})=\sqrt{\varepsilon_\beta^2({\bf k})+|\Delta_{ss^\prime}^{(\beta)}({\bf k})|^2}$ is the energy of Bogoliubov quasiparticle. Putting Eq. (\ref{Eq:Bogolibov1}-\ref{Eq:uvvk}) into Eq. \ref{Eq:gap}, one can derive the linear gap equations which read
\bea
\Delta^{\beta}({\bf k})=\frac{1}{A}\sum_{\beta^\prime,{\bf k}^\prime}\chi^{\beta,\beta^\prime}({\bf k},{\bf k}^\prime)\Delta^{\beta^\prime}({\bf k}^\prime),\label{Eq:gapequations}
\eea
with
\bea
&& \chi_{{\bf k},{\bf k}^\prime}^{(\beta,\beta^\prime)}=g^{\beta,\beta^\prime}({\bf k},{\bf k}^\prime)\frac{\tanh\left[\frac{\varepsilon_{\beta^\prime}({\bf k^\prime})}{2k_BT}\right]}{2\varepsilon_{\beta^\prime}({\bf k^\prime})}, \label{Eq:chi}
\eea
near the superconducting transition temperature $T_c$. The superconducting transition temperature $T_c$ is the highest temperature such that the matrix $\chi_{{\bf k},{\bf k}^\prime}^{(\beta,\beta^\prime)}$ yields eigenvalue 1 in the  discrete momentum and band spanned space\cite{chou2021acoustic}.
\begin{figure}
	\includegraphics[width=0.5\textwidth]{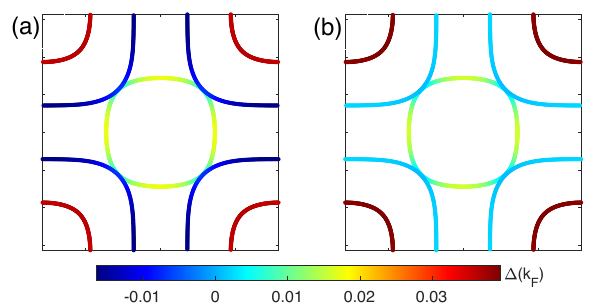}
	\caption{The pairing gap with fixed interlayer coupling $g_2^f=0.12$ while varying intralayer coupling $g_1^f=0.006$ (a) and $0.072$ (b) in the full-coupling case with chemical potential $\mu=0$. }
	\label{fig:Fig4}
\end{figure}

\section{Superconducting state for the full-coupling case}\label{sec:superconducting_state_for_the_full_coupling_case}
% We first focus on the full-coupling case shown in Eq. (\ref{Eq:full}).
We begin by systematically analyzing the full-couling case defined in Eq. \ref{Eq:full}. In the band space, the interacting strength $g_1^{\beta,\beta^\prime}({\bf k},{\bf k}^\prime)$ from the contribution of intralayer coupling is 
\bea
&&g_1^{\beta,\beta^\prime}({{\bf k},{\bf k}^\prime})=g_1^f\sum_{\eta\alpha}U^*_{\eta\alpha \beta}({\bf k})U^*_{\eta\alpha \beta}(-{\bf k})U_{\eta\alpha \beta^\prime}(-{\bf k}^\prime)\nonumber \\
&&\times U_{\eta\alpha \beta^\prime}({\bf k}^\prime),  
\eea
and $g_2^{\beta,\beta^\prime}({\bf k},{\bf k}^\prime)$ contributed by the interlayer coupling is 
\bea
&&g_2^{\beta,\beta^\prime}({{\bf k},{\bf k}^\prime})=g_2^f\sum_{\alpha}U^*_{t\alpha \beta}({\bf k})U^*_{b\alpha \beta}(-{\bf k})U_{b\alpha \beta^\prime}(-{\bf k}^\prime) \nonumber \\
&&\times U_{t\alpha \beta^\prime}({\bf k}^\prime). 
\eea
After putting $g^{\beta,\beta^\prime}({\bf k},{\bf k}^\prime)=g_1^{\beta,\beta^\prime}({{\bf k},{\bf k}^\prime})+g_2^{\beta,\beta^\prime}({{\bf k},{\bf k}^\prime})$ into Eq. (\ref{Eq:chi}), we can numerically obtain the pairing gap by solving Eq .(\ref{Eq:gap}) and (\ref{Eq:chi}). 

To clearly understand the role of intralayer and interlayer coupling in the pairing symmetry, we separately examine the intralayer and interlayer couplings and plot the resulting pairing symmetries for various doping levels in Fig.~\ref{fig:Fig2}.
% As shown in Fig. \ref{fig:Fig1} (a), the $T_c$ follows the same behavior as the DOS when varying the doping $\mu$. Near each VHS doping, $T_c$ reaches a local maximum value and it has the order $10$ K. 
% Fig.~\ref{fig:Fig2}(a-c) reveal that the pairing symmetry induced by the interlayer BCS coupling in the full-coupling case shows that the pairing gaps in pockets $\alpha$ and $\gamma$ have the same sign, while the gap between pockets $\alpha$ and $\beta$ has the opposite sign, with fixed parameters $g_1^f=0$ and $g_2^f=0.12$ for several doping conditions. 
Fig.~\ref{fig:Fig2}(d-f) show that the pairing symmetry induced by the interlayer BCS coupling in the full-coupling case results in pairing gaps with the same sign in the $\alpha$ and $\gamma$ pockets, while the gap between the $\alpha$ and $\beta$ pockets has the opposite sign, for fixed parameters $g_1^f=0$ and $g_2^f=0.12$ under various doping conditions. Thus, our numerical calculations for the full-coupling case with only the interlayer coupling indicate that the superconducting state hosts an $s\pm$-wave pairing symmetry. \begin{figure}
	\includegraphics[width=0.5\textwidth]{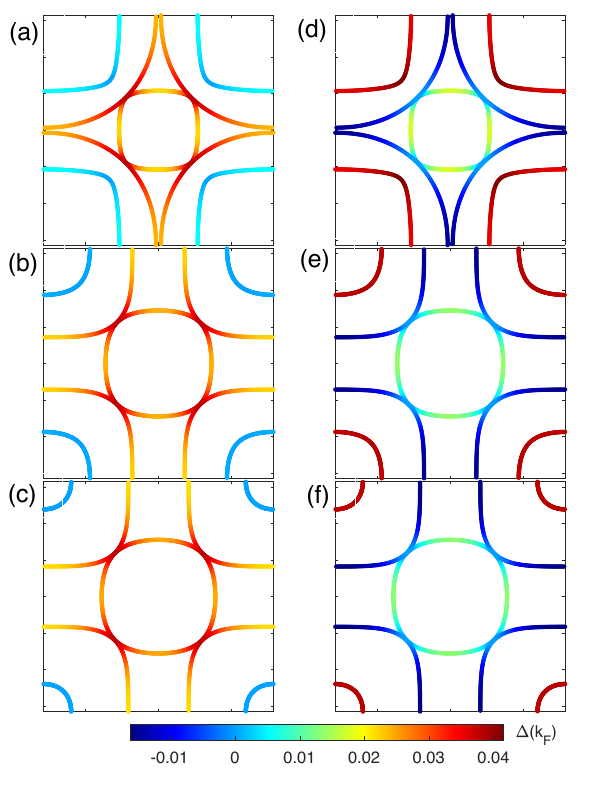}
	\caption{The pairing gap in the half-coupling case with different parameters. (a), (b) and 
 (c) illustrates the numerical result with only intralayer coupling $g_1^f=0.18$ for $\mu=-0.328eV$, $0eV$ and $0.1eV$ respectively. (d), (e) and (f) illustrates the numerical result with only interlayer coupling $g_2^f=0.18$ for $\mu=-0.328eV$, $0eV$ and $0.1eV$ respectively.}
	\label{fig:Fig5}
\end{figure}
% Fig. \ref{fig:Fig2}(a-c) reveal that the pairing symmetry induced by the intralayer BCS coupling in the full-coupling case demonstrate that pairing gaps in pockets $\alpha$ and $\gamma$ have the same sign but a opposite sign between $\alpha$ and $\beta$.  Thus, our numerical calculations in the full-coupling case with only the intralayer coupling reveal the superconducting state host the $s\pm$ wave pairing symmetry. 
Furthermore, to address the origin of the $s\pm$ pairing symmetry induced by interlayer coupling, we plot the pairing interactions in the band space in Fig. \ref{fig:Fig3}(a-c). As illustrate in Fig. \ref{fig:Fig3}(a,b), the  pairing interaction $g^{\alpha,\alpha}$ 
 and $g^{\alpha\gamma}$ indicate negative Josephson coupling both within $\alpha$ intra-Fermi pocket and between $\alpha$ and $\gamma$ Fermi pockets, intriguing to pairing gaps with the same sign within the same Fermi pockets and between $\alpha$ and $\gamma$ Fermi pockets. Meanwhile, the negative  $g^{\alpha,\beta}$ will favor the opposite sign pairing gap between $\alpha$ and $\gamma$ Fermi pockets. Conversely, the pure intralayer coupling scenario will sustains the conventional $s++$-wave superconducting state with uniform pairing gap across all pockets as shown in Fig. \ref{fig:Fig2}(a-c) with fixed parameter $g_1^f=0.12$ and $g_2^f=0$.

The competition between these two pairing channels manifests a superconducting phase transition in the ($g_1^f$, $g_2^f$) parameter space. Fig. \ref{fig:Fig4} explicitly demonstrates that increasing the intralayer coupling strength 
 $g_1^f$ drives a phase transition from the $s\pm$-wave to $s++$-wave superconducting ground state, marked by the suppression of inter-pocket gap sign reversal.
% Combined the pure interlayer and intralayer coupling scenarios,  we can expect that the competition between interlayer and intralayer couplings will drive a superconducting phase transition in the parameter space $(g_1^f, g_2^f)$ as shown in Fig. \ref{fig:Fig4} where increasing intralayer coupling will drive a phase transition from the $s\pm$ to $s++$ superconducting state. 

\begin{figure}
	\includegraphics[width=0.48\textwidth]{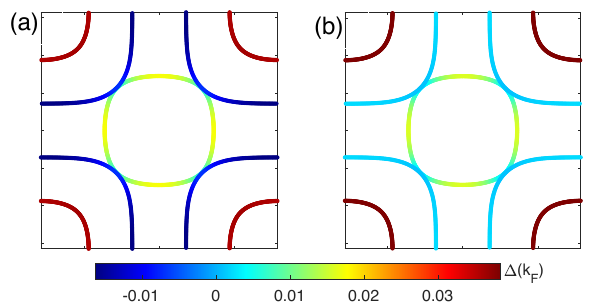}
	\caption{The pairing gap in the half-coupling case with fixed interlayer coupling $g^{h}_2 =0.18$  while varying intralayer coupling $g_1^h = 0.016$ (a) and $0.05$ (b) with chemical potential $\mu=0$}
	\label{fig:Fig6}
\end{figure}

\section{Pairing symmetry in half-coupling case} \label{sec:pairing_symmetry_in_half_coupling_case}
In this section, we further investigate the superconducting pairing symmetry in half-coupling case which is demonstrated in Eq. \ref{Eq:half}. In this case the interaction strength $g^{\beta,\beta^\prime}$ is followed by

\bea
&&g_1^{\beta,\beta^\prime}({{\bf k},{\bf k}^\prime})=g_1^h\sum_{\eta}U^*_{\eta x^2 \beta}({\bf k})U^*_{\eta x^2 \beta}(-{\bf k})U_{\eta x^2 \beta^\prime}(-{\bf k}^\prime)\nonumber \\
&&\times U_{\eta x^2 \beta^\prime}({\bf k}^\prime),  
\eea
and $g_2^{\beta,\beta^\prime}({\bf k},{\bf k}^\prime)$ contributed by the interlayer coupling is 
\bea
&&g_2^{\beta,\beta^\prime}({{\bf k},{\bf k}^\prime})=g_2^h U^*_{t z^2 \beta}({\bf k})U^*_{b z^2 \beta}(-{\bf k})U_{b z^2 \beta^\prime}(-{\bf k}^\prime) U_{t\alpha \beta^\prime}({\bf k}^\prime). \nonumber \\
\eea
where $g_{1,2}^h$ is the intralayer or interlayer pairing strength in the half coupling case as mentioned before. 

Following the methodology employed in the full-coupling scenario, we systematically investigate the individual effects of intralayer and interlayer couplings. By numerically solving the self-consistent equations Eq. (\ref{Eq:gap}) and (\ref{Eq:chi}) with the interaction $g^{\beta,\beta^\prime}$, we plot the pairing gaps in Fig.~\ref{fig:Fig5}. As shown in Fig. \ref{fig:Fig5}(d-f) with fixed parameters $g_1^h=0$ and $g_2^h=0.18$ with several doping conditions, the interlayer coupling also favors the $s\pm$-wave pairing symmetry as the same as in the full-coupling case. In Fig. \ref{fig:Fig5}(a-c), we use the different intralayer coupling strengths $g_2^h$ to drive the superconducting state and obtain the pairing gap under different doping conditions. Then we reach the same conclusion in the half-coupling case as in the full-coupling scenario, the intralayer coupling $g_2^h$ also tends to induce the conventional $s++$-wave pairing symmetry, due to the same reasons as discussed in the full-coupling case.

% As the same in the full-coupling case, we also first consider the consequence of intralayer and interlayer coupling, respectively. Performing the numerical calculations for Eq. (\ref{Eq:gap}) and (\ref{Eq:chi}) after inserting $g^{\beta,\beta^\prime}$, we plot the pairing symmetry in Fig. \ref{fig:Fig4} with given \textcolor{red}{$g_1^h=?$ and $g_2^h=0$}. As shown in \textcolor{red}{Fig. \ref{fig:Fig4}(a-d)}, the intralayer superconducting interaction also lead to a $s\pm$ superconducting pairing symmetry. The reason also can be attributed to the repulsive Josephson coupling $-g^{\beta,\gamma}$, and attractive Josephson coupling $-g^{\alpha,\gamma}$. As the same discussion in the full-coupling case, the pairing symmetry in the half-coupling case induced by inter-layer coupling is also the $s$-wave.

% Here we emphasis that the pairing symmetry considered in  both the full-coupling and half-coupling cases emerge from the nature of electronic band structure of the pressured La$_3$Ni$_2$O$_7$.
\begin{figure}
	\includegraphics[width=0.48\textwidth]{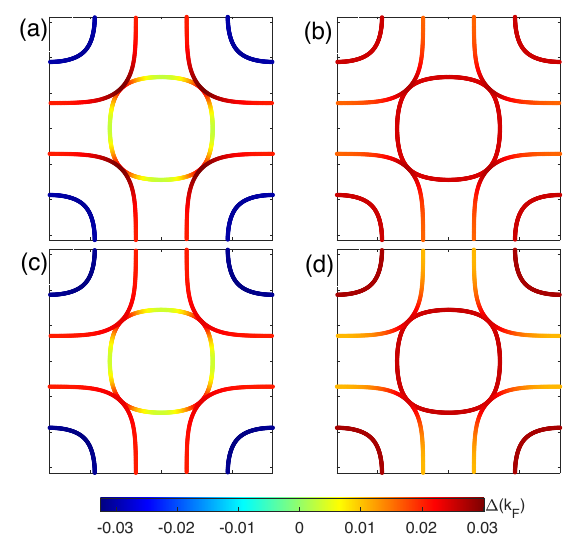}
	\caption{The pairing gaps with fixed $\mu=0eV$. (a) and (b) marks the pairing gap in the full-coupling case with fixed parameters $g_1^f=0.006$ and $g_2^f=0.12$ while varying pair hoping $g_P=-0.1$ and $0.1$, respectively. (c) and (d) illustrates the pairing gap in the half-coupling case with fixed parameters $g_1^h=0.016$ and $g_2^h=0.18$ while varying pair hoping $g_P=-0.1$ and $0.1$, respectively.}
	\label{fig:Fig7}
\end{figure}

\section{The role of pair hopping interaction} \label{sec:the_role_of_pair_hopping_interaction}
In multiband superconductivity, a pertinent question arises regarding the role of the pair-hopping term. To address this, we investigate the pairing symmetry by incorporating the pair-hopping term in both full and half-coupling scenarios. The pair hopping term reads
% In multiband superconductivity there will arise a general question what kind of the role the pair hoping term plays in. To address this question, in this section we investigate the pairing symmetry after including the pair-hopping term in both full and half-coupling cases. 
\bea
H_P=&&-g_P\sum_{\eta;{\bf k},{\bf k}^\prime;s,s^\prime}c^{\dagger}_{\eta x^2 s}(-{\bf k})c^{\dagger}_{\eta x^2 s^\prime}({\bf k})c_{\eta z^2 s^\prime}({\bf k}^\prime)\nonumber \\
&&\times c_{\eta z^2 s}(-{\bf k}^\prime)
\eea
where $g_P$ is the pair hopping strength. In the band space, it can be rewritten as
\bea
&&H_{P}=-\sum_{{\bf k},{\bf k}^\prime}\sum_{\beta \beta^\prime;ss^{\prime}}g_P^{\beta,\beta^\prime}({\bf k},{\bf k}^\prime) c_{\beta s}^\dagger({\bf k})c_{\beta s^\prime}^\dagger(-{\bf k})\nonumber \\
&&\times c_{\beta^\prime s^\prime}(-{\bf k}^\prime)c_{\beta^\prime s}({\bf k}^\prime)
\eea
with 
\bea
&&g_P^{\beta,\beta^\prime}({{\bf k},{\bf k}^\prime})=g_P\sum_{\eta}U^*_{\eta x^2 \beta}({\bf k})U^*_{\eta z^2 \beta}(-{\bf k})U_{\eta x^2 \beta^\prime}(-{\bf k}^\prime)\nonumber \\
&&\times U_{\eta z^2 \beta^\prime}({\bf k}^\prime). 
\eea
Then we can obtain 
\bea
g^{\beta,\beta^\prime}=g_1^{\beta,\beta^\prime}({{\bf k},{\bf k}^\prime})+g_2^{\beta,\beta^\prime}({{\bf k},{\bf k}^\prime})+g_P^{\beta,\beta^\prime}({{\bf k},{\bf k}^\prime}).
\eea
Compared to Fig.\ref{fig:Fig4}(a), the numerical results presented in Fig.\ref{fig:Fig7}(a) in the full-coupling model indicate that a negative pair-hopping interaction $g_P$ tends to drive the superconducting state toward an alternative $s\pm$-wave state. In this state, the pairing gap on the $\alpha$ Fermi surface shares the same sign as the gap on the $\beta$ Fermi surface but has the opposite sign compared to the gap on the $\gamma$ Fermi surface.  In contrast, a positive $g_P$ in the full-coupling case favors the conventional $s++$-wave pairing symmetry. The preference for $s\pm$-wave symmetry by negative and $s++$-wave symmetry by positive $g_P$ can be attributed to the fact that the  Fermi surface is primarily composed of $d_{3z^2-r^2}$  orbital weight, while both the $\alpha$ and $\beta$ Fermi surfaces have significant contributions from the $d_{x^2-y^2}$ orbital. Fig. \ref{fig:Fig7} (c) and (d) lead to the conclusion that in the half-coupling case the pair-hopping strength $g_P$ plays a similar role as in the full-coupling case.

\section{Discussions and Conclusions}\label{conclusion}\label{sec:discussions_and_conclusions}
Our numerical results demonstrate that two coupling constant both need to exceed a finite critical value to get superconducting state (see Appendix B). To our best knowledge, the largest value of EPC constant $\lambda$ in LNO is estimated to be 1.75 under a pressure of 22.4 GPa\cite{talantsev2024debye}. The EPC constant, in combination with a typical Coulomb pseudopotential of $\mu^*\simeq 0.1$, is sufficient to generate a superconducting state within the McMillan framework. Furthermore, according to DFT calculations\cite{zhan2024cooperation}, the dominant contributions to the phonon spectrum arise from two out-of-plane $A_{1g}$ modes, which are associated with interlayer electron–phonon coupling. Assuming, for simplicity, that these two modes contribute equally to the $\lambda$, the interlayer coupling $g_2$ can be estimated as $g_2/A\sim \lambda/\rho(0)/A\simeq 0.6 eV$ where $\rho(0)\simeq 10 eV^-1 nm^{-2}$ is the density of state at $\mu=0$. This estimated $g_2/A$ exceeds the typical interlayer antiferromagnetic exchange interaction $J_\perp\sim 0.32eV$\cite{Ku2024PRL}, and can be comparable to or even surpass the effective suppression from the onsite Hubbard $U$ when projected onto the interlayer pairing channel. Therefore, an interlayer phonon-mediated attraction strong enough to dominate over the residual repulsion is physically plausible.

In summary, we have systematically investigated the superconducting pairing symmetry mediated by effective attractive interactions in the celebrated high-temperature superconductor material La$_3$Ni$_2$O$_7$ under pressure. We considered two models, namely the full-coupling case in which the Ni-$d_{x^2-y^2}$ orbital and $d_{3z^2-r^2}$ orbital are treated equally in the interlayer and intralayer interactions, and the half-coupling case where the interlayer and intralayer coupling only includes $d_{3z^2-r^2}$ and $d_{x^2-y^2}$ orbital, respectively. By solving the linear gap equation numerically, we have found that in the both cases the interlayer coupling tends to trigger the $s\pm$ superconducting state with a sign change between $\beta$ and $\alpha$ ($\gamma$) Fermi surfaces, while the intralayer coupling leads to a pure $s++$-wave without sign change between Fermi surfaces. Hence our calculations have revealed the competitive effect between interlayer and intralayer coupling, and also uncover that the final pairing symmetry arise from the nature of electronic wave-functions near Fermi level of the material under pressure. In addition, we also have considered the pair hopping interaction between $d_{3z^2-r^2}$ and $d_{x^2-y^2}$ orbitals, and have demonstrated that the repulsive pair hopping favors the $s\pm$ pairing symmetry as expected due to the fact $\gamma$ and $\alpha$ Fermi surface mainly is mainly contributed from the $d_{3z^2-r^2}$ and $d_{x^2-y^2}$, respectively.

\section{Acknowledge}
We thank Xianxin Wu for very helpful discussions. Xinloong Han is supported by National Natural Science Foundation of China (Grant No. 12404162). Yucong Yin and Boyang Liu are supported by the National Science Foundation of China (Grant No. NSFC-11874002).

\section{Appendix}
\subsection{Numerical details}
\begin{figure}[ht]
	\includegraphics[width=0.45\textwidth]{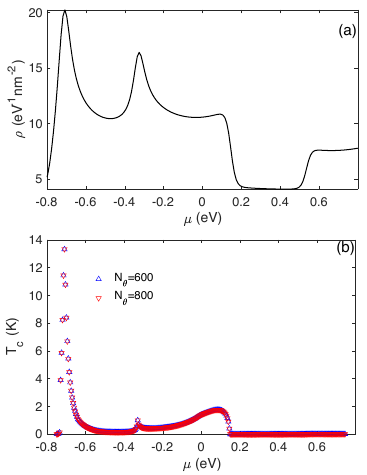}
	\caption{The density of state $\rho$ (a) and the transition temperature $T_c$ (b) as a function of chemical potential $\mu$. In (b), the blue upward triangles and red downward triangles correspond to calculations using $N_\theta=600$ and $N_\theta=800$ in the half-coupling case, respectively. (b) is calculated with parameters $g_1^h=0$, $g_2^h=0.18$ and $g_P=0$.}
	\label{fig:sFig1}
\end{figure}

To numerically solve the gap equations Eq. (\ref{Eq:gapequations}) in the main text, we firstly rewrite it in the polar coordination and only consider momentum near the Fermi surfaces,
\bea
\Delta^{\beta}(\theta)=\sum_{\beta^\prime}\int_0^{2\pi}\frac{d\theta^{\prime}}{(2\pi)^2} \chi_{\theta,\theta^\prime}^{\beta,\beta^\prime}\Delta^{\beta^\prime}(\theta^\prime),
\eea
with
\bea
\chi_{\theta,\theta^\prime}^{\beta,\beta^\prime}=g^{\beta,\beta^\prime}(\theta,\theta^\prime)\int_{|\varepsilon_{\beta^\prime}(k_{\theta^\prime})|<\Lambda}k_{\theta^\prime} dk_{\theta^\prime}\frac{\tanh\left[\frac{\varepsilon_{\beta^\prime}({\bf k^\prime})}{2k_BT}\right]}{2\varepsilon_{\beta^\prime}({\bf k^\prime})}.
\eea
where $\Lambda$ is the energy cutoff. In this paper, we set $\Lambda=0.15eV$. We also take the equiangular division with $N_\theta$ pieces. In Fig. \ref{fig:sFig1}(b), In Fig. 8(b), the numerical results obtained with $N_\theta=600$ and $N_\theta=800$ are nearly identical, indicating convergence. Therefore, we choose $N_\theta=600$, unless otherwise specified in this paper.

\begin{figure}[ht]
	\includegraphics[width=0.46\textwidth]{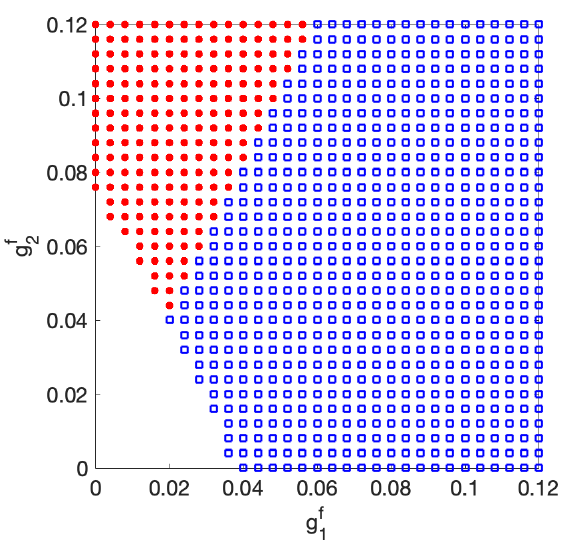}
	\caption{The phase diagram for the full-coupling case with $g_P=0$. The blue square denote the conventional s-wave pairing symmetry while the red dot represents the $s\pm$-wave pairing symmetry. The blank (or white) region indicates that no superconductivity emerges within our lowest accessible temperature range.}
	\label{fig:sFig4}
\end{figure}

\begin{figure}[ht]
	\includegraphics[width=0.48\textwidth]{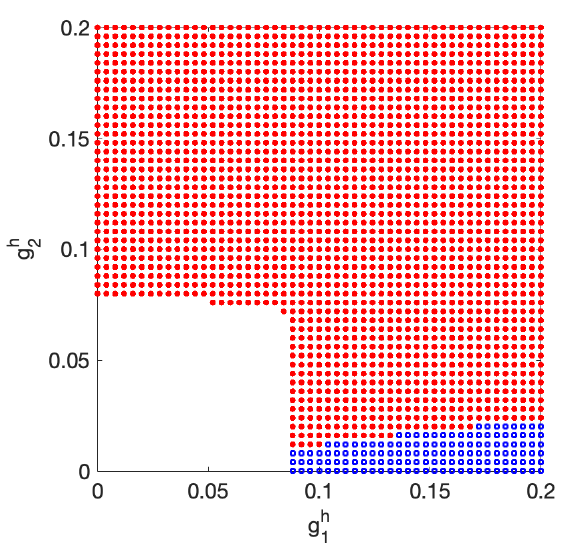}
	\caption{The phase diagram for the half-coupling case with $g_P=0$. The blue square denote the conventional s-wave pairing symmetry while the red dot represents the $s\pm$-wave pairing symmetry.}
	\label{fig:sFig5}
\end{figure}
\subsection{Phase diagrams}
To get a better understanding how intralayer and interlayer coupling compete in both cases, in this section we calculate the phase diagrams for the full-coupling and half-coupling cases. As illustrated in Fig. (\ref{fig:sFig4}), The phase diagram reveals that $s$-wave superconductivity dominates a wide region of parameter space, while the $s\pm$-wave superconductivity phase only appears in a relatively narrow region where the interlayer interaction is significantly weaker than the intralayer interaction. Conversely, $s\pm$-wave superconductivity in the phase diagram shown in Fig. \ref{fig:sFig5} for the half-coupling case, occupies the vast majority of the parameter space, demonstrating its robust nature. From Fig. (\ref{fig:sFig4}) and (\ref{fig:sFig5}), we also notice that the interlayer and intralayer coupling strength in both cases must exceed their critical values to get the superconducting state.

\bibliographystyle{apsrev4-2}
%\bibliography{reference}

\end{document}